\documentclass[]{article}

\usepackage[T1]{fontenc}
\usepackage{graphicx}
\usepackage{hyperref}
\usepackage{authblk} 
\usepackage{abstract}

\begin{document}

\title{Towards a Distributed Platform for Normative Reasoning and Value Alignment in Multi-Agent Systems}

\author[1]{Miguel Garcia-Bohigues}
\author[1]{Carmengelys Cordova}
\author[1]{Joaquin Taverner}
\author[1]{Javier Palanca}
\author[1]{Elena del Val}
\author[1]{Estefania Argente}

\affil[1]{Universitat Politécnica de València - VRAIN}
\affil[ ]{\{migarbo1, ccorgar3, joataap, japaca1, edelval, eargente\}@vrain.upv.es}
\affil[ ]{\url{https://vrain.upv.es/}}

\maketitle
\begin{abstract}
This paper presents an extended version of the SPADE platform, which aims to empower intelligent agent systems with normative reasoning and value alignment capabilities. Normative reasoning involves evaluating social norms and their impact on decision-making, while value alignment ensures agents' actions are in line with desired principles and ethical guidelines. The extended platform equips agents with normative awareness and reasoning capabilities based on deontic logic, allowing them to assess the appropriateness of their actions and make informed decisions. By integrating normative reasoning and value alignment, the platform enhances agents' social intelligence and promotes responsible and ethical behaviors in complex environments.
\end{abstract}


\section{Introduction}

In recent years, the field of artificial intelligence has witnessed significant advancements in the development of intelligent agent systems. These systems aim to emulate human-like decision-making processes and behavior in order to interact effectively in complex and dynamic environments. A crucial aspect of human decision-making is the consideration of norms, 
which are social rules that govern acceptable behavior within a group or society.  Norms play a fundamental role in guiding individuals' actions and interactions, influencing their choices, and ensuring social order and cohesion. Another fundamental aspect of today's society is values, as they influence how people interact with each other, make decisions, and address both social and cultural issues. Values refer to the fundamental beliefs and principles that guide individuals and societies in determining what is considered important, desirable, and morally right. Values inform decision-making and influence attitudes, behaviors, and priorities, shaping the way individuals perceive the world, interact with others, and make choices in various contexts.

The lack of ethical and moral values can lead to inappropriate behaviors and the violation of people's rights, which can have negative consequences in society. On the other hand, values can have a positive impact, as they foster responsible and supportive behaviors, promote tolerance and respect for others, and contribute to the development of a just and equitable society.


To create more socially intelligent agent systems, researchers have recognized the need to incorporate both normative reasoning capabilities and ethical and moral values into agent models \cite{lorini2021logic}. Normative reasoning involves the evaluation of norms and their impact on decision-making, taking into account factors such as the rewards and/or penalties associated with norm compliance or violation. By integrating normative reasoning, agents can assess the appropriateness of their actions within the context of established norms, enhancing their ability to interact effectively in social environments \cite{boman1999norms}.

Values play a crucial role in shaping the behavior and decision-making processes of autonomous agents. In MAS, values can be seen as internal constructs that represent an agent's preferences, priorities, and moral foundations, providing a framework for agents to evaluate and prioritize their actions, interactions, and goals \cite{bench2017norms}.
By incorporating values into the decision-making process, agents can align their actions with desired principles, ethical guidelines, and social norms \cite{lorini2015logic}.

Endowing actors with the ability to reason about norms and values could improve not only their security but also their trustworthiness. Indeed, value alignment has become one of the fundamental principles that should govern actors and is an important part of responsible AI \cite{sierra2021value}.

One of the key challenges in the current landscape of Multi-Agent Systems (MAS) platforms is the absence of normative reasoning capabilities, which limits the agents' ability to effectively evaluate and comply with social norms. Normative reasoning plays a vital role in simulating realistic and socially intelligent agent behavior, enabling agents to assess the appropriateness of their actions within the context of established norms \cite{mahmoud2014review,broersen2001boid}. Another prevalent issue is the lack of real-time distributed functionality in existing MAS platforms, hindering the seamless coordination and communication among agents operating in dynamic and distributed environments. Additionally, the current platforms often lack efficient mechanisms for managing external agent connections, impeding the integration and interaction of agents across different systems and networks. These limitations pose significant obstacles to the development of robust and scalable MAS applications and call for the exploration of novel approaches that address these issues and provide a more comprehensive platform for building intelligent agent systems.

In this paper, we present an extension of the SPADE (Smart Python Agent Development) platform, a widely used framework for developing intelligent agent systems. This extension enables the development of agents capable of reasoning about norms. We introduce a new framework facilitating the development of norms based on deontic logic, encompassing concepts like prohibition, permission, and obligation. Our framework equips agents with normative awareness and normative reasoning capabilities. When an agent intends to perform an action, the normative reasoning process evaluates the existing norms in the environment and informs the agent of potential sanctions or rewards associated with the action. The agent then employs this information to make a decision on whether to proceed with the action or not.


The rest of the paper is organized as follows:  Section~\ref{sec:related-work} provides an overview of the relevant literature related to the scope of this study;
Section~\ref{sec:proposal} presents the proposed approach in detail; and Section~\ref{sec:conclusions} concludes the paper by summarizing the main findings, discussing implications, and suggesting avenues for future research and development.


\section{Related work}\label{sec:related-work}
 In this section, we present a brief overview of the existing literature and research in three key areas: values, norms, and multi-agent systems (MAS) platforms. These areas are crucial for understanding the foundation and context of our work.
 
\subsection{Values}
There are different theoretical and philosophical appreciations of the concept of value, such as that of 
Shein \cite{schein2010organizational}, who conceptualizes values as "the reasons given to explain the way things are done"; Rokeach \cite{rokeach1973nature}, defines the concept of value as those "beliefs" that people hold about desirable end states and/or behaviors, and which therefore transcend concrete situations by guiding the selection and evaluation of situations and behaviors; or Azjen \cite{ajzen1980understanding} who expresses that values are the objects, ideas or beliefs that are appreciated and that affect the way of looking at things, observing aspects such as vital, ethical, pleasant and useful.

The most widely accepted value system is that of Schwartz \cite{schwartz2012overview}, where values are understood as broad motivational goals that transcend a single situation or action and that serve as the criteria to evaluate them. Also, Schwartz states that across all societies, the same 10 values can be observed: self-direction, stimulation, hedonism, achievement, power, security, conformity, tradition, benevolence, and universalism. It is the order of preference between them that makes this society different and not the lack or presence of some of them.
Equipping agents with the ability to reason about norms and values could improve not only the safety but also the trustworthiness of these agents. Values are what we find important in life and can be used, for example, in explanations of agents' behavior \cite{winikoff2018bad}.
Moreover, values principal use is to model and control the behavior of the members of a society by taking the role of an internal guide that restricts their actions or by stating some obligations within the society in certain cases\cite{schwartz2012overview}.

\subsection {Norms}
Norms are guidelines or rules established by authorities, institutions, or communities to regulate behavior within a group or society \cite{o2017kinds}. These rules define what is considered acceptable in a given context and outline the consequences of compliance or violation. Norms can take the form of formalized laws, regulations, codes, or unwritten rules transmitted among members of a social group or inferred from observed behavior. 

Norms have been used in Artificial Intelligence (AI) research with the aim of regulating the life of software entities and the interactions between them. Specifically, norms have been proposed in the field of AI to address coordination and security issues in multi-agent systems (MAS), as well as to model legal issues in electronic institutions and electronic commerce \cite{criado2011open}. The most promising application of MAS technology is its use to support the use of AI to support the development and deployment of software entities and support open distributed systems. 
Open distributed systems are characterized by heterogeneity of participants, untrustworthy members, conflicting individual goals, and a high possibility of non-compliance with specifications. The main characteristic of agents in these systems is autonomy. It is this autonomy that requires regulation, and norms are a solution to this requirement. In such systems, problems are solved through cooperation between various software agents. The norms prescribe what is allowed, forbidden, and obligatory in societies. Thus, they define the benefits and responsibilities of the members of the society, and, consequently, agents can plan their actions according to their expected behavior \cite{criado2011open}.

There are different classifications of norms, such as those proposed by Tuomela \cite{tuomela1995importance}, Dignum \cite{dignum1999autonomous}, Boella \cite{boella2008substantive}, Criado \cite{criado2011open}, Mahmoud \cite{mahmoud2014review} or Savarimuthu \cite{savarimuthu2011norm}. However, from all these proposals, four main types of norms \cite{argente2020normative} can be distinguished according to the entity that promulgates them or the audience to which they are addressed: (i) institutional norms established by authorities such as government or company management, (ii) social norms or conventions that emerge from repeated interactions within a group or society, (iii) interaction norms affecting specific groups for limited periods (e.g., "legal contracts" or "formal agreements"), and (iv) private norms that individuals impose on themselves \cite{lliguin2018challenges,sterelny2019norms}.

For our work, we consider that norms and values are strictly interrelated. We have identified value and normative implications in three of the four norm types described above.

First, private norms are deeply rooted in an individual's values, as they reflect their personal understanding of what is right, just, and morally acceptable. Values provide the foundation upon which private norms are built, shaping an individual's perception of appropriate conduct and influencing their choices and actions. Therefore, private norms and values have a reciprocal relationship, as values inform the development of private norms, while private norms serve as manifestations of an individual's underlying values. Together, private norms and values play a crucial role in shaping an individual's moral compass and guiding their interactions with others and their engagement in society.
Second, social norms can be understood as the formal reflection of the values inside the society. Social norms, as stated before, are norms that are derived from the behavior of the different agents that interact in that space. Also, these kinds of norms ensure that these agents can coexist in this environment avoiding undesired situations. 
Finally, institutional norms are related to values in two ways. On one hand, the social norms can be seen as a formal representation of the institutional values. On the other hand, this kind of norm is also related to the agent values, being reinforced if aligned with them, or rejected if there's a conflict between the norms and the agent values.

Recent research has explored ways to integrate values and norms into practical reasoning. For example, Mercuur et al. \cite{mercuur2019value} have incorporated values and norms into social simulations. In their work, agents may act according to values or norms, but they do not consider the interaction between norms and values. However, several authors have argued that agents should use value-based arguments to decide what action to take,
including whether to comply with or violate the norms \cite{bench2017norms,serramia2018exploiting}. Cranefield et al. \cite{cranefield2017no} have studied
how to consider values in the plan selection algorithm used by a BDI\footnote{BDI (Belief-Desire-Intention) is a popular model for the development of intentional agents, in which agents are endowed with a set of mental attitudes. In this model, agents are able to infer knowledge and reason about internal states and changes that occur in the environment. This reasoning enables the agent to perform actions in order to achieve its goals.} agent, choosing the plan that is most consistent with the agent's values to achieve a given goal. However, other aspects of value-based reasoning are not considered, such as the interaction between values, goals, and norms. Values and norms play a more fundamental role in the functioning of a BDI agent, and a combination of these two mental attitudes allows agents to behave in a way that is more aligned with human expectations \cite{szabo2020understanding}.

\subsection{MAS platforms}
A multi-agent system platform is a software framework that provides tools, libraries, and infrastructure for developing, deploying, and executing multi-agent systems (MAS). Typically, a platform provides a set of features and functionalities that simplify the development, coordination, communication, and management of multiple autonomous agents within a system. Over the years, numerous platforms have been proposed to support multi-agent systems. One of the best known is JADE (Java Agent Development Framework) \cite{bellifemine2001jade}. JADE provides an open-source framework compatible with FIPA (Foundation for Intelligent Physical Agents) with communication-based on ACL (Agent Communication Language). Similarly, in \cite{bordini2005jason} Jason is proposed. Jason is a programming language derived from the language for BDI (Belief-Desires-Intentions) \cite{rao1997modeling} agents AgentSpeak \cite{rao1996agentspeak}. Jason also provides the language interpreter and a platform for agent development. That platform manages both the environment and the lifecycle and communication of the agents. 

Another interesting approach using Python programming language is SPADE \cite{palanca2020spade}. SPADE is a middleware for multi-agent systems that represents an evolution of the traditional multi-agent system platforms by means of incorporating a careful selection of concepts and modern technologies in the areas of distributed systems, instant messaging, asynchronous systems, and open systems. The agent model that SPADE uses is based on:
\begin{itemize}
\renewcommand\labelitemi{--}
    \item \textit{A connection mechanism}: by which each agent registers in SPADE by using a unique identifier (JID).
    \item \textit{Behaviors}: which are independent tasks that execute the agent’s actions. Behaviors can be of several types: Cyclic, One-Shot, Periodic, Time-Out, and Finite State Machine. Each one is designed to support a typical execution requirement
    \item \textit{A message dispatcher}: which SPADE associates with each registered agent. This component acts as a mailman, redirecting any incoming message to the particular behavior(s) that may be expecting it, and relaying the outgoing messages to the SPADE’s communication system. 
\end{itemize}

\section{Proposal}\label{sec:proposal}


In this section, we present a novel framework that allows the development of norms in multi-agent systems using deontic logic\footnote{The code corresponding to the current version of this framework can be found at the author's GitHub: \url{https://github.com/javipalanca/spade_norms}}, which encompasses concepts such as prohibition, permission, and obligation. 

Our framework extends the SPADE platform to support the development of agents with normative awareness and normative reasoning capabilities. As SPADE is a generic platform for distributed multi-agent systems, it provides support for the interconnection of agents developed in different languages (e.g., Python or Java) and with different architectures (e.g., finite state machine, or BDI). This makes it difficult to standardize internal protocols for norm representation and normative reasoning. For example, for an agent based on finite state machines, a norm can be expressed by a transition between states, while in a BDI model, a norm can be expressed as a belief. Therefore, in this paper, we propose the use of what we call a \textit{normative backpack}. When the agent first registers in the system, the environment provides it with a personal \textit{normative backpack}. This \textit{normative backpack} is used as an add-on that is associated with the agent and mediates between the multi-agent system environment and the agent, facilitating norm normalization and normative reasoning. Thus, when the agent intends to perform an action, it requests information about the norms from the \textit{normative backpack}. The \textit{normative backpack} then executes a normative reasoning process, evaluating the existing norms in the environment, and informs the agent of the possible penalties or rewards of performing the action. Finally, the agent will then use this information to decide whether or not to perform the action.

We model the use of norms within the SPADE platform through agent organizations. Each organization can have its own internal norms, roles, domains, and hierarchies. In this way, agents can join one or more organizations depending on their needs and objectives. When an agent joins an organization, it is provided with the organization's \textit{normative backpack} and is informed about the set of actions allowed within that organization. 

Within organizations, roles play a fundamental aspect in allowing the definition of different typologies of agents, depending on their individual characteristics or the specific function they perform in the context of the organization. This makes it possible to establish a framework of behavior and coordination between agents, promoting cooperation and facilitating the decision-making process. On the other hand, the definition of the domain of norms and roles in organizations allows the establishment of clusters that significantly reduce the computational cost of the normative reasoning process.


\subsection{Norm specification}\label{sec:norm-specification}

One of the fundamental aspects in the development of a system capable of reasoning about norms is the formal and semantic specification of the norms.  It should also be taken into account that, in general, platforms for multi-agent systems (and especially SPADE) are designed to allow the interconnection of embedded systems or IoT devices. In such systems, performance, effectiveness, and efficiency are critical. In addition, as mentioned above, different agents developed in different programming languages and architectures coexist on the SPADE platform. Therefore, the formal specification of a normative model designed for this kind of platform must guarantee accessibility, efficiency, and flexibility. 

Taking these requirements into consideration, we propose the use of a generic semantic specification that can be adapted to most normative scenarios. In our model a norm is defined by the tuple $\langle id, t, c, ac, r, p, rs, d, inv, issu \rangle$ where:

\begin{itemize}
\renewcommand\labelitemi{--}
    \item $id$: is a unique identifier of the norm. It is usually a string describing the name of the norm or even an identification number. 
    \item $t$: corresponds to the deontic type of the norm. Currently only the deontic operators ``prohibition'' or ``permission'' are considered.
    \item $c$: represents the activation condition of the norm. This condition can be defined by a logical expression or a pointer to a predefined function in the system environment. 
    \item $ac$: is an internal flag that allows the system to identify when a norm is active. Like the condition, it can be a logical expression or a pointer to a function. It can be omitted, and the system will then assume that the norm is always active.
    \item $r$: corresponds to the reward that the agent will receive when complying with a norm.
    \item $p$: corresponds to the penalty that the agent will receive for breaking the norm.
    \item $rs$: is the set of roles affected by the norm. It is also optional and if not provided it is assumed that the norm affects all agents in the system.
    \item $d$: is the domain of the norm. Its main purpose is to facilitate the computational design of the system by creating different categories (domains) of both actions and norms and grouping them together. As with roles, norms with a specific domain will only affect actions in a specific domain. As this can also be omitted, if no domain is provided, one will be assumed by default and will affect all actions.
    \item $inv$: is a flag indicating whether the norm can be violated or not. Norms classified as inviolable should have priority in the agent's decision-making process. In our work, we consider private norms to be generally inviolable, as they reflect the agent's values. For this reason, the default value is $True$ although this can be modified by the expert in the domain. 
    \item $issu$: is a label that identifies the source of the norm. This label can take the values: ``Self'', when it is a private norm or a concern (through which ethical and moral values are reflected); ``Society'', if it is a social norm; or ``organization'', if the norm comes from an organizational authority or regulator. 
\end{itemize}

To see how this fits into a real example, we select the taxi station scenario presented in \cite{argente2022NEA}. This scenario simulates a taxi station with a two-lane queue. Taxi drivers must wait the entire queue before taking any customers. Once they are at the first position of this queue, they can pick up customers if the whole group fits in the taxi. If not, the taxi driver must return to the last position in the queue using the second line. Apart from that, drivers are not allowed to work more than 8 hours so a 30-minute break is mandatory when the time comes.

In this example, we could identify different types of norms. For example, as institutional norms we could define "Taxi drivers must respect work regulations, such as not exceeding a limit of 8 hours of continuous work" or "Taxi drivers must take a mandatory 30-minute break when the limit of working hours is met"; "Taxi drivers cannot exceed the taxi capacity when picking up customers". As social norms, we could have: "Taxi drivers must wait in the queue and follow the order of arrival to pick up customers". 
As private norms, we could have norms related to the values of security and tradition ("Taxi drivers are prohibited from violating customers' privacy and disclosing information provided during journeys"); to the values of safety and conformity ("Taxi drivers are prohibited from violating traffic rules and driving in an unsafe manner at any time"), or to the values of benevolence and universalism ("Taxi drivers are prohibited from being rude or disrespectful to customers, providing friendly and helpful treatment at all times").

For the sake of simplicity, two examples of norms are given below. The first one, of an institutional type, indicates that: "Taxi drivers must wait in the designated queue and follow the order of arrival to pick up customers". In other words, \textbf{``a taxi driver is forbidden to jump the queue''}, and its formal representation will be:
\begin{verbatim}
{   
    id: "respectLine",
    t: PROHIBITION, 
    c: driverQueuePos == numTaxisQueue, 
    ac: True,
    r: 0, 
    p: -1,
    rs: [DRIVER],
    d: QUEUE,
    inv: False, 
    issu: ORGANIZATION 
}
\end{verbatim}

The second norm, also of an institutional nature, states that \textbf{``it is forbidden for a taxi driver to carry more customers than there are seats in his car''}, and could be formally represented as:

\begin{verbatim}
{   
    id: "respectCapacity",
    t: PROHIBITION, 
    c: taxiCapacity >= NumClientsWaiting, 
    ac: True,
    r: 0, 
    p: -5,
    rs: [DRIVER],
    d: PICKING,
    inv: True, 
    issu: ORGANIZATION 
}
\end{verbatim}

We could go so on with the other deductible norms that follow the same structure, but for the sake of simplicity, we will specify only these two.

\subsection{Normative backpack}
As stated in the introduction of this section, the main objective of this framework is to develop a component for SPADE in such a way that it can have broad normative support for many different MAS solutions. Typically, in SPADE, the agent's available actions and decision-making process are specified within the agent's behavior. Note that here by behavior we refer to the SPADE conception of behavior\footnote{SPADE available behaviors and its explanation can be found in \url{https://spade-mas.readthedocs.io/en/latest/behaviours.html}}. Also note that actions are domain-dependent and so are norms, however, our solution must be domain-free so that it can be used in any context. 

To address this problem, we propose the development of an external component that is dynamically attached to the agent and gives it normative reasoning capabilities. This component is what we have called the ``normative backpack''. It can be understood as a backpack that is given to the agents once they enter the organization. We consider an agent to enter an organization whether it is created within a closed organization or whether an agent joins an open organization. In both cases, the agent will be given the normative extension. 

In there, the agent will have at hand all the norms that have been predefined at the organization as well as its concerns (private norms that may have appeared dynamically due to interactions with the environment or with other members of the organization, allowing the agent to align itself with the values of the society). In addition, the agent will also have here the actions it can perform. As we have commented above, SPADE agents have their actions encoded in behaviors. Therefore, we force the actions to be moved from the behaviors to the normative component. This way we can guarantee that before performing any kind of action, the system will first handle the normative consequences. Of course, this is left to the developer, who must specify which actions will be regulated and which will not, by adding them to the backpack.

The agent will also have the Normative Engine, a powerful inference method in charge of deciding, for a given coded action, whether or not there is a normative problem with it. This is done by checking both the norms of the organization and the agent's own concerns as a single set of norms. 
To do this, the Normative Engine will filter both entities by domain and role. The purpose of this filtering is to maximize the efficiency of the system by checking only those norms that affect the specific domain or role. 

The Normative Engine will return as a result all the information that the agent may need to make a decision on whether or not to comply with the norms. That is the regulatory status of the action (ALLOWED, FORBIDDEN, INVIOLABLE or NOT\_REGULATED), the norms that allow and forbid the action (if any in both cases), and the total reward and penalty obtained in case of performing and not performing the action.

To handle this decision process, the plugin also provides the development of a custom instance of a SPADE agent with a specific component developed for the normative context. This is the Normative Reasoning Engine, which is a customized inference engine that is responsible for deciding precisely whether or not a norm or list of norms is met. Since it is highly dependent on the domain and the specific task to be solved, it can be overridden and customized by the developer as desired.

\subsection{Normative action process}
So far we have discussed both the formal representation of the norm and the components of the normative backpack. The next step is to see how they all coexist and work together to provide agents with a normative framework. To do this, we will use the norms formalized above as an example. It is important to say that currently, the framework allows either Prohibition norms or Permission norms, but not both at the same time. This means that we can have two types of systems:
\begin{itemize}
    \item In the case of prohibition norms, we will assume an environment in which everything is  \textbf{permitted} except what a norm explicitly prohibits. 
    \item In the case of permission norms, we will assume the opposite. In these cases, everything will be \textbf{forbidden} except the explicitly permitted actions and situations.
\end{itemize}

\begin{figure}[ht!]
    \centering
    \includegraphics[trim = 68mm 45mm 90mm 9mm, clip, width=\linewidth]{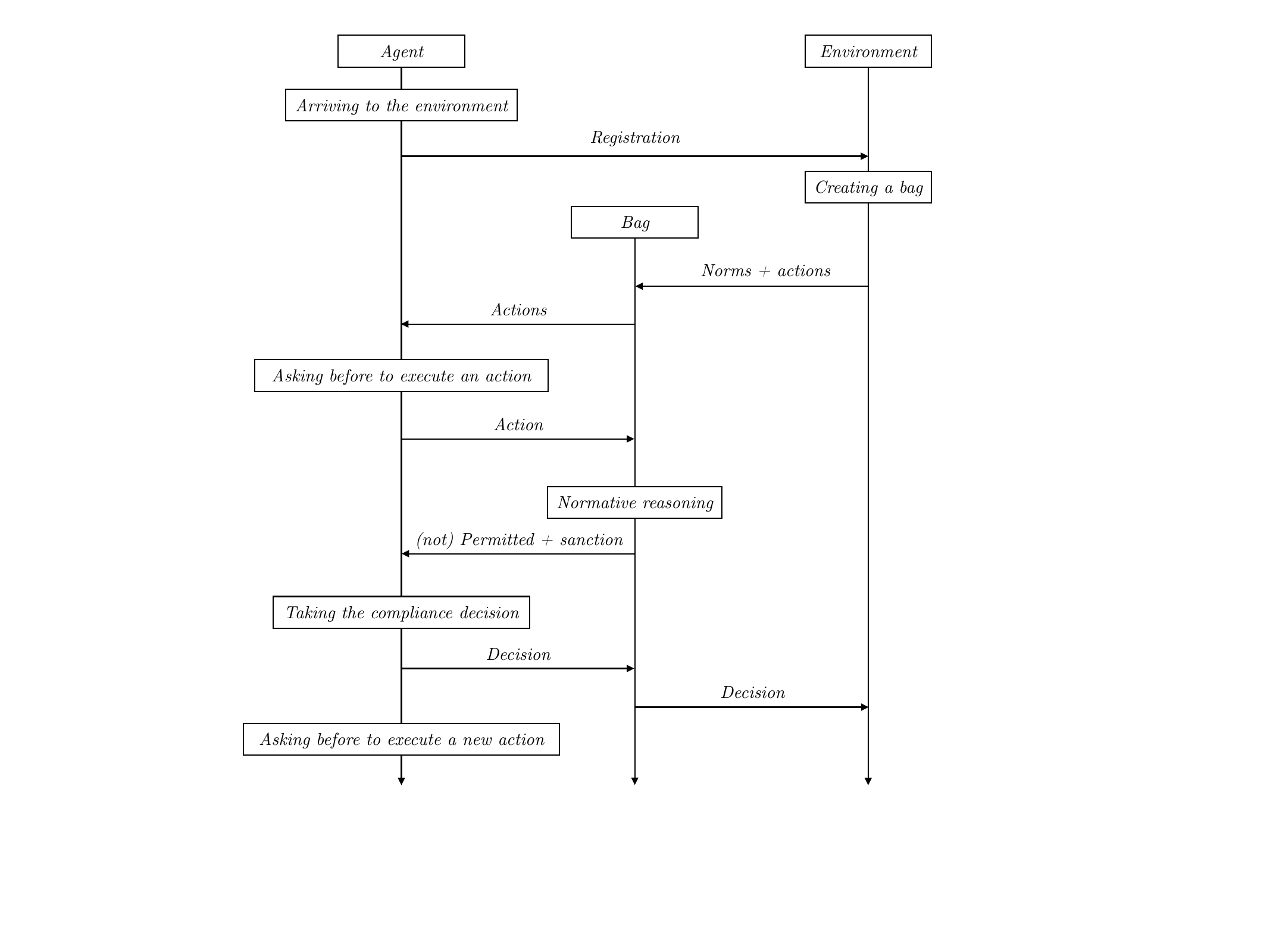}
    \caption{Workflow of the Normative Action Process }
    \label{fig:NAP-alg}
\end{figure}

Note that the norms that we have used as examples are prohibitions, so our example system will be a prohibition scenario. So, first of all, we have to define which normative actions we have. For the two example norms we can find the \textit{Queue} action and the \textit{PickClients} action. Then, the first step will be to add them to the normative backpack. After that, as developers we have to add the norms that control the states we want our agents to avoid (like picking up more people than there are seats in the car). Figure \ref{fig:NAP-alg} shows the flowchart of the normative action process for a general-purpose scenario.  

As we have seen, once the agent enters the system, it does not matter if it has joined the organization or if it has been recently created, it receives the normative backpack which has been previously filled with both norms and actions that are regulated within the organization. At this point, the agent will behave as usual until it decides to perform a regulated action. That is an action that is marked inside the backpack as normative. At this point, the agent, before performing the action, asks the backpack for the normative implication of performing that action at that specific moment with the parameters desired by the agent. The backpack then launches the normative engine determines whether the action is allowed or not and returns the previously detailed information to the agent. It is then that the agent reasons about whether or not it pays off to perform the action. In either case, it notifies both the backpack and the environment. 

This algorithm transferred to our example, will behave as follows: Our agent has arrived at the first position in the queue where there are six people waiting to get a taxi. Before telling them to get into the taxi, the driver checks the norms. Norms tell him that he has only 4 seats available and, by picking up those six people, he will be breaking an inviolable norm, so he decides to leave the queue without picking up the customers (he avoids performing the action). Nevertheless, as he returns to the end of the queue, he sees that a colleague has picked the six customers and that there is now a new group of only three people waiting to be chosen. At this point, he sets out to perform the Queue action. As his goal is to get as much money as possible, he checks the feasibility of skipping the whole queue and picking up the customers directly. To do so, he checks again the current rules and sees that he will break a norm and that this will mean a penalty of -1.  Knowing this, the agent proceeds to see if the reward obtained for picking up the clients compensates him for breaking the norms. 
As he will obtain 2 points for each customer he picks, he decides to jump the queue, break the law, and pick the customers directly.

\section{Conclusions and future work}\label{sec:conclusions}
In this article, we provide an overview of different types of norms and theories of values. We emphasize the importance of norms and normative reasoning in Multi-Agent Systems (MAS) and highlight the close relationship between values and norms, showcasing how the latter can be used to model the former. Furthermore, we present a formal representation of norms that enables the construction of normative and value systems. We discuss key aspects of the normative algorithm, including the backpack and the normative engine. 

While the platform currently supports functional normative reasoning, there is still much work to be done to fully incorporate value reasoning. Presently, the platform only accommodates either prohibition or permission norms within the same organization, without allowing for their combination. Additionally, the current approach uses norms to define both agent and society values, but we recognize that values extend beyond merely shaping behavior and should be considered as goal producers. Incorporating values in this manner is an area of ongoing research. Lastly, we believe that BDI agents, along with reinforcement learning agents, can greatly benefit from such platforms, and thus, there is a need for direct and customized development tailored to these types of agents.



\subsubsection*{Acknowledgements} This work is partially supported by Spanish Government by projects PID2020-113416RB-I00, PRE2021-098964 and TED2021-131295B-C32.

\end{document}